\documentclass[jcp,twocolumn,amsmath,amssymb,showpacs]{revtex4}

\usepackage[usenames,dvipsnames]{color} 
\usepackage{graphicx}
\definecolor{LinkColor}{rgb}{0,0,.5}
\usepackage[colorlinks=true,linktoc=Red,linkcolor=Red,citecolor=LinkColor,urlcolor=LinkColor]{hyperref}
\renewcommand{\emph}{\textit}
\newcommand\Id{\leavevmode\hbox{\small1\normalsize\kern-.33em1}}
\newcommand{\sz}{\sigma_z}
\newcommand{\ket}[1]{\left\vert{#1}\right\rangle}
\newcommand{\ham}{{\mathcal{H}}}
\newcommand{\tr}[1]{\textrm{Tr}\left\{{#1}\right\}}

\graphicspath{{Figure2submit/}}

\begin{document}

\title{Initialization and Readout of Spin Chains for Quantum Information Transport}

\author{Gurneet Kaur} 
\author{Paola Cappellaro}\email{pcappell@mit.edu} \affiliation{Department of Nuclear
Science and Engineering, Massachusetts Institute of Technology,
Cambridge, Massachusetts 02139, USA}

\begin{abstract}
Linear chains of spins acting as quantum wires are a promising approach to achieve  scalable quantum information processors. Nuclear spins in  apatite crystals provide an ideal test-bed for the experimental study of quantum information transport, as they closely  emulate a one-dimensional spin chain.
Nuclear Magnetic Resonance techniques can be used to drive the spin chain dynamics and probe the accompanying transport mechanisms.
Here we demonstrate initialization and readout capabilities in these spin chains, even in the absence of single-spin addressability. These control schemes enable preparing  desired states  for quantum information transport and probing their evolution under the transport Hamiltonian. We further optimize the control schemes by a detailed analysis of $^{19}$F NMR lineshape.
\end{abstract}

\pacs {03.67.Hk, 03.67.Lx, 75.10.Pq, 76.90.+d}

\maketitle


\section{Introduction}

Control  over small quantum systems and the ability to perform simple quantum algorithms have been demonstrated on a variety of physical systems ranging from trapped ions~\cite{Cirac95} and electrons~\cite{Ciaramicoli03}, to neutral atoms and molecules in optical lattices~\cite{Monroe02}, to superconducting circuits~\cite{Vion02} and semiconductor quantum dots~\cite{Petta05}, to nuclear and electronic spins~\cite{Chuang98,Cory00,Cappellaro09}. Although algorithms involving more than one qubit have been executed~\cite{Vandersypen01,Gulde03,Hodges07,Chiaverini05,DiCarlo09,Schindler11}, a vital requirement for a quantum computer --\textit{scalability while preserving fidelity}-- has not yet been achieved in any physical system. The use of linear chains of spins as quantum wires  to couple basic memory units is a promising approach to address this issue~\cite{Campbell08,Jiang07}. These spin chains have the ability to transmit quantum information via the free evolution of the spins under their mutual interaction~\cite{Bose03,Christandl04,Christandl05,Cappellaro07l,Kay07,Kay10}.
While advances in fabrication techniques have made physical implementation of spin wires possible~\cite{Weis08,Toyli10,Naydenov10,Spinicelli11},  the level of precision available is not yet adequate. Therefore, natural systems such as crystals where spin position is precisely set by nature are a preferred choice for exploring such applications.

Owing  to their unique geometry, nuclear spin systems in apatite crystals have emerged as a rich test-bed to probe quasi-one-dimensional (1D) dynamics, including transport and decoherence~\cite{Cappellaro07a,Zhang09,RufeilFiori09,Ramanathan11}. The crystal wherein $^{19}$F (or $^1$H) nuclei are aligned along one axis, emulates a collection of 1D chains. The dynamics of these spin chains have been studied by various nuclear magnetic resonance (NMR) techniques~\cite{Sur75,Cho93,Cho96}.
In  our previous work~\cite{Cappellaro07l,Cappellaro11}, we have shown that the natural dipolar interaction among the spins can be manipulated via the available collective control to simulate the Hamiltonian driving quantum transport. The lack of single-spin addressability in this ensemble systems however would seem to prevent  creating and measuring a single-spin excitation as required to study transport.
Still, we demonstrated experimentally~\cite{Cappellaro07a,Zhang09} in the Fluorapatite (FAp) system that one can prepare the spin system in an initial state in which polarization is localized at the ends of the spin chain, a state that well simulates the conditions for spin-excitation transport~\cite{Cappellaro11}.

In this paper we take   further steps toward enabling the experimental study of quantum transport in spin chains: we introduce an experimental technique to read out  the spins at the chain extremities and we show how to prepare a two-spin encoded state that is able to transfer quantum information.
We use these initialization and readout techniques to study the dynamics under the transport Hamiltonian (the Double Quantum (DQ) Hamiltonian~\cite{Munowitz87,Ramanathan11}).
Additionally, we probe the spin chain dynamics by creation and evolution of multiple quantum coherences~\cite{Ramanathan03}, which  present well-characterized state dependent signatures.  We use both these techniques to demonstrate preparation as well as readout of spins at the chain ends, thus verifying  important required tasks towards simulation of quantum information transfer. We further validate the addressability of ends of the chains by a detailed analysis of $^{19}$F lineshape in FAp.

These techniques will make it possible to explore errors affecting the transport fidelities as well as control schemes to mitigate them, in an experimental setting, where the interactions among spins are not limited to the ones tractable by solvable models.

\section{Transport in mixed-state spin chains}
\subsection{Spin chain dynamics}
\label{sec:transport}
Linear chains of spin-1/2 particles have been proposed as quantum wires to transport quantum information between distant nodes of a distributed quantum computing architecture. Transport can occur even in the absence of individual control of the chain spins, as it is mediated by the spins mutual interactions. In the most studied model, energy conserving spin flip-flops (mediated by the isotropic XY Hamiltonian) drive the transport of a single-spin excitation~\cite{Bose03,Christandl04,Christandl05,Cappellaro07l,Kay07,Kay10}. This model has been recently extended to the case where the initial state of the chain cannot be perfectly controlled, and thus it is found in a mixed-state rather than its ground state~\cite{Fitzsimons06,DiFranco08c,Markiewicz09,Yao11,Cappellaro11}.

Spin chains that are in a maximally mixed-state are particularly interesting from the point of view of experimental study of quantum information transfer. This state, corresponding to the infinite temperature, can be easily achieved experimentally and has been shown to provide a direct simulation of pure state transport~\cite{Cappellaro11}. Additionally, extension to mixed-state chain enables using the so called double quantum (DQ) Hamiltonian:
\begin{eqnarray}
{\cal H}_{DQ} &= &\sum_{j <\ell} \frac{b_{j\ell}}{2}
(\sigma^x_j\sigma^x_{\ell}-\sigma^y_j\sigma^y_{\ell}),
\label{eq:dqHam}
\end{eqnarray}
to drive transport, although it does not conserve the number of spin excitations. This Hamiltonian can be easily obtained from the natural dipolar Hamiltonian with only collective control~\cite{Cappellaro07l,Cappellaro11} and it is related to the isotropic XY Hamiltonian (which instead cannot be generated from the dipolar interaction) via a similarity transformation. The extension to mixed-states and to  DQ Hamiltonian open the possibility to study experimentally quantum information transport in nuclear spin chains with NMR techniques.
Under our experimental conditions  (strong external magnetic field, B$_0$=7T and room temperature), the  initial equilibrium state is the Zeeman thermal state,
\begin{eqnarray}
\rho'_{th}(0) &\propto & \exp(-\varepsilon \sigma^z)\approx \openone -
\varepsilon \sigma^z,
\label{eq:thermalstate}
\end{eqnarray}
where $ \sigma^z= \sum_{j}\sigma^z_j $ and $ \epsilon = \gamma B_{0}/k_{B}T$.
Since the identity does not evolve and does not contribute to the signal, we will focus on the deviation of the density operator from the maximally mixed state, $\delta\rho\sim\rho-\Id$.
In the absence of individual spin addressing, transport within a chain can be studied by preparing a polarization excess at one end of the chain, that is, a state  where one spin at the chain extremities is polarized while the remaining spins are fully mixed, $\delta\rho\sim\sz^1$.
Because of the symmetry between the two chain ends, the state we can prepare experimentally~\cite{Cappellaro07a,Zhang09}, which we call  ``end-polarized state'', is given by
\begin{eqnarray}
\delta\rho_{\textrm{end}}(0) &=& \sigma^z_1+\sigma^z_N.
\label{eq:endpolstate}
\end{eqnarray}
The end-polarized state simulates the dynamics of a single-spin excitation in a pure-state spin chain. This state  can  transfer  a bit of classical information by encoding it in the sign of the polarization. This encoding is, however, not enough to transfer quantum information, which requires additional transfer of information about the phase coherence of a state.  A two-qubit encoding allows for the transport of a bit of quantum information~\cite{Markiewicz09,Cappellaro11}. For transport via the DQ Hamiltonian, this encoding is given by the basis $\ket{0}^{\textsf{dq}}_L=\ket{00}$ and $\ket{1}^{\textsf{dq}}_L=\ket{11}$. The operator basis for transport via mixed states under DQ Hamiltonian is thus given by:
    \begin{equation}
  	\begin{array}{ll}
      \sigma_{xL}^{\textsf{dq}}=\frac{\sigma_x^1\sigma_x^2-\sigma_y^1\sigma_y^2}2,\
  	\ &\ \
  	\sigma_{yL}^{\textsf{dq}}=\frac{\sigma_y^1\sigma_x^2+\sigma_x^1\sigma_y^2}2,\ \  	
     \\ \sigma_{zL}^{\textsf{dq}}=\frac{\sigma_z^1+\sigma_z^2}2,\ \ &\ \
  	\Id_L^{\textsf{dq}}=\frac{\Id+\sigma_z^1\sigma_z^2}2 .
  \end{array}
  \label{eq:LogDQ}
  \end{equation}
Starting from any of the above  initial states, the evolution  under the DQ Hamiltonian directly simulates the transport dynamics within a chain.

In the limit of nearest neighbor (NN) coupling only, the evolution under DQ Hamiltonian is exactly solvable by invoking a Jordan-Wigner mapping onto a system of free fermions~\cite{Lieb66,Feldman96,Ramanathan11}. The resulting dynamics of the observables we analyzed in our experiments have been reported in the literature (see e.g. \cite{Ramanathan11,Cappellaro11}) and are reviewed in Appendix \ref{sec:App_Analytical} for completeness.
Isolated, linear spin chains with NN couplings  is an accurate model for the experiments, given the experimental timescales used~\cite{Zhang09}.  Comparison of the theoretical model with the experimental results thus allows us to validate our initialization and  readout methods.

To gather more insights into the states generated by the evolution under the DQ Hamiltonian, we experimentally measured  multi-spin correlations via multiple quantum NMR experiments. This experimental technique (see   Appendix  \ref{sec:App_MQCNMR}) allows measurement of multi-spin state dynamics by indirectly encoding their signatures into single spin terms. The  dynamics of   quantum coherence intensities can as well be calculated analytically in the limit of NN couplings~\cite{Feldman96,Cappellaro07l,Cappellaro11} and we  review these results in Appendix  \ref{sec:App_MQCNMR}.

  \subsection{Preparing and reading out desired spin states}
  \label{sec:EndChain}
  To probe the quantum transport dynamics it is necessary to prepare the spins at the ends of the chain in a perturbed state (such as $\delta\rho_{\textrm{end}}$), which is then left to evolve under the transport Hamiltonian. Furthermore, to observe the transport, measurement of the end chain spins would be desirable. Unfortunately, in a system of dipolarly coupled homonuclear spins (such as FAp) it is not possible to achieve frequency or spatial addressability of individual spins. Still, here we show that we can approximate these preparation and readout steps with the control at hand, combining unitary and incoherent spin manipulation. The key observation is that even in the absence of frequency addressability, the dynamics of the end-chain spins under the internal dipolar Hamiltonian is  different from the rest of the spins in the chain. Indeed, the  spins at the ends of the chain are coupled to only one nearest neighbor whereas spins in the rest of the chain have two nearest neighbors.  This fact can be exploited to experimentally prepare the spins at the chain ends in a desired state~\cite{Cappellaro07a,Zhang09} as well as subsequently read out  this state as explained below.

  When the initial thermal equilibrium state is rotated to the transverse plane by a $\pi$/2 pulse, we create a state $\sum_{k=1}^{N}\sigma^k_x$ which evolves under the internal dipolar Hamiltonian. Due to fewer numbers of couplings with neighboring spins, the spins at the end of the chain have slower dynamics as compared to the rest of the chain. Thus, one can select a particular time $t_1$ such that whereas the state of the spins at the ends is still mainly $\sigma_{x}$, the  rest of the spins have evolved to many-body correlations. A second $\pi/2$ pulse brings the magnetization of the end spins back to the longitudinal axis while an appropriate phase cycling scheme cancels out other terms in the state. We used the following pulse sequence and appropriate phase cycling scheme to select the ends of the chains,
  \begin{equation}\label{seq:P1}
      \pi/2|_\alpha \text{ --- } t_1 \text{ --- } \pi/2|_{-\alpha},
       \tag{P1}
       \nonumber
  \end{equation}
  with $\alpha=$x,y to average out terms that do not commute with the total z-magnetization.
  For FAp crystals we found that the optimal $t_1$ time  (which we will call ``end-selection time'') is given by 30.3 $\mu$s~\cite{Cappellaro07a,Zhang09} . Further details on how we optimized this time are given in Section~\ref{sec:lineshape}.

The end-selection scheme forms the basis for a strategy to  prepare other states, presented in Eq. \ref{eq:LogDQ},  required for quantum information transport.  
  In order to prepare these encoded states experimentally, we use the following scheme. We first prepare the end polarized state
  $\delta\rho_{\textrm{end}}(0) = \sigma^z_1+\sigma^z_N$
   and then let the system evolve under DQ Hamiltonian for a short time. Applying a double quantum filter by an appropriate phase cycling scheme selects the desired state:
  $\delta\rho_{yL}\propto\left.\sigma_{yL}^{\textsf{dq}}\right|_{1,2}+\left.\sigma_{yL}^{\textsf{dq}}\right|_{n-1,n}$.
  Similarly, a $\pi/4$ collective rotation around $z$, prior to the double-quantum filter, is needed to select the $\delta\rho_x^L$ operator.

  A similar combination of unitary and incoherent spin control can be used to read out the spins at the end of the chain. In inductively measured NMR, the observable is the collective magnetization of the spin ensemble. To simulate the readout of a different observable, the desired state must be prepared prior to acquisition. The sequence \ref{seq:P1} can be used for this purpose:  we call this the ``end readout step''.
  We note that the sequence  used for readout cannot in general be a simple inversion of the end-selection step (which is not a unitary --reversible-- operation). A proper phase cycling should ensure that the state prior to the end-selection sequence  has contributions mainly from population terms ($\propto \sz^k$). This property is verified for the states produced from evolution under the DQ Hamiltonian in  1D systems, making   the \ref{seq:P1} sequence effective for the end-readout step; a more complex phase cycling would be needed for more general states.


\section{Experimental Methods}
\label{sec:mqc}
\begin{figure*}[t]
 \includegraphics{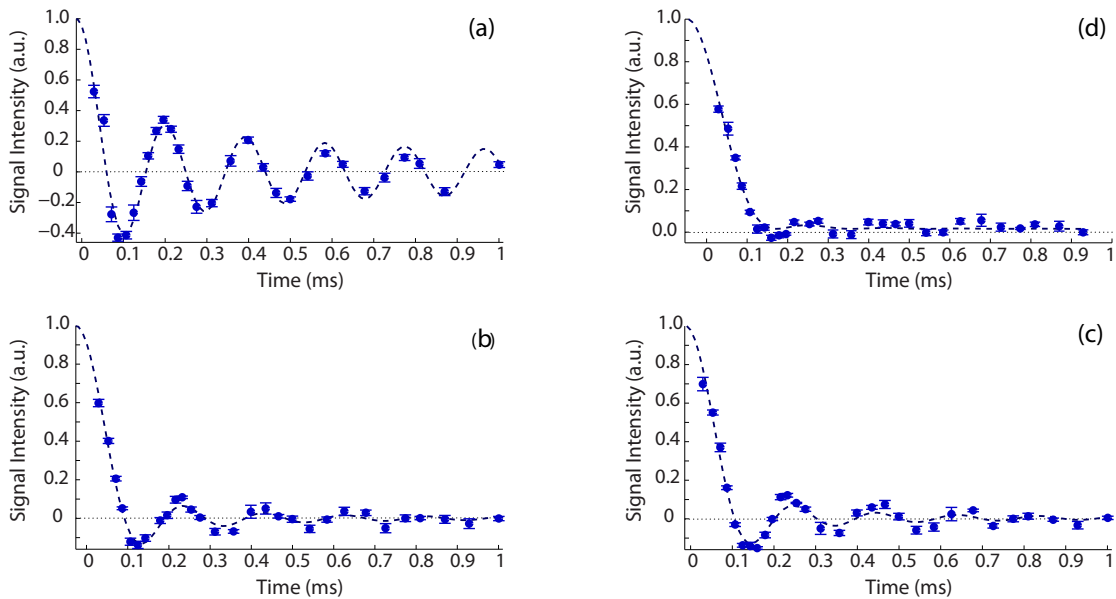}
   \caption{(Color online) Evolution under the DQ Hamiltonian: (a) Initial state: $\delta\rho_{th}$. Readout: collective magnetization.  (b) Initial state: $\delta\rho_{\textrm{end}}$. Readout: collective magnetization.  (c) Initial state: $\delta\rho_{th}$. Readout: end readout.
   (d) Initial state: $\delta\rho_{\textrm{end}}$. Readout: end readout.
   Data points are the experimental data and  lines are the  fits  using the analytical model described in appendix~\ref{sec:App_Analytical}.  Error bars are given by the offset of the signal from zero.  The fitting gives the following values for the dipolar coupling: 8.165 (a), 8.172(b), 8.048 (c) and 8.63 (d) $\times10^3$rad/s. }
    \label{fig:gkDQall}
  \end{figure*}

Experiments were performed in a 7~T widebore magnet with a 300 MHz Bruker Avance Spectrometer and a probe tuned to 282.4MHz for $^{19}$F measurement.  A pure, single crystal of Fluorapatite [Ca$_5$(PO$_4$)$_3$F] grown by flux method  was used for the measurements~\cite{Oishi94,Teshima11}. High purity of the crystal is confirmed by long relaxation time (T$_1$=1100~s) of $^{19}$F spins. FAp crystals have a hexagonal geometry with space group P63/m. The $^{19}$F nuclei form linear chains along the $c$ axis, each surrounded by 6 other chains. The intra-nuclear spacing within a single chain is d=0.3442nm and chains are separated by D=0.9367nm. When placed in a strong magnetic field, the nuclear spins interact via the secular dipolar Hamiltonian,
\begin{equation}
\ham_{\textsf{dip}}=\sum_{j< l}^n b_{j l}
\left[\sigma_z^j\sigma_z^{ l} - \frac1{2}
(\sigma_x^j\sigma_x^{ l}+\sigma_y^j\sigma_y^{ l})\right],
\label{eq:dipHam}
\end{equation}
where the couplings depend on the relative positions as $b_{jl}=(\mu_0/16\pi) (\gamma^2 \hbar /r_{j l}^3) (1-3\cos^2\theta_{j l})$, with $\mu_0$ being the standard magnetic constant, $\gamma$ the gyromagnetic ratio, $r_{j l}$ the distance between nucleus $j$ and $l$, and $\theta_{j l}$ the angle between $\vec r_{j l}$ and the $z$-axis, respectively. Due to 1/r$^{3}$ dependence of dipolar coupling, the largest ratio of in-chain to cross chain coupling is close to 40. For our experiments, the crystal was aligned parallel to the external magnetic field in order to maximize this ratio. It has been  shown that under these settings, and for short evolution times, couplings across different chains can be neglected and the system can be considered as a collection of 1D chains~\cite{Zhang09}.

We performed two sets of experiments for each of the different initial states and readouts.
First we probed the transport dynamics driven by the DQ Hamiltonian. For this purpose, the collective or end-chain magnetization was measured as we increased the evolution time $t$ under the DQ Hamiltonian. We used a standard 8-pulse sequence~\cite{Yen83} to implement the DQ Hamiltonian. The length of the $\pi$/2 pulse  was $1.45\mu$s. The evolution time was incremented by varying inter-pulse delay from $1\mu$s to $6.2\mu$s and the number of loops was increased from 1 to 12.  A recycle delay of 3000~s was used for these measurements.
We restricted the evolution to a timescale where the ideal model applies and errors arising from leakage to other chains and next-nearest neighbor couplings are small~\cite{Zhang09}. In this timescale, the initial perturbation travels across $\approx17$ spins~\cite{Ramanathan11}, a distance much shorter than the  average  length of the  chains. Thus, within the timescale used in this work, only the polarization starting from one end of the chain and moving away towards the other end was observed: polarization reaching the other end could not be observed.

In the second set of experiments, we let the initial state evolve under the DQ Hamiltonian and  measured the multiple quantum coherences  (MQC) to gather more information on the evolved state (see Appendix~\ref{sec:App_MQCNMR} for details on the experimental method).
The inter-pulse delay was varied from 1$\mu$s to 6$\mu$s and the number of loops was increased from 1 to 3. We encoded coherences up to order 4 with a $K=4$ step phase cycling.
Since these measurements involve phase cycling, a shorter recycle delay of 1000s could be used.
\subsection{Experimental Results: Spin Transport}
Fig. \ref{fig:gkDQall}(a)  shows the observed evolution of the collective magnetization under the DQ Hamiltonian starting from the thermal initial state (Eq. \ref{eq:thermalstate}). The data points were fitted with the analytical function (Eq. \ref{eq:DQ}) described in appendix \ref{sec:App_Analytical}.
In figure~\ref{fig:gkDQall}(b) we plot the system dynamics when starting from an end-polarized state, Eq. \ref{eq:endpolstate} (where polarization is localized at the ends of the chain) and reading out the collective magnetization.  Figure ~\ref{fig:gkDQall}(c) shows a complementary measurement where we start from thermal initial state,  given by the collective magnetization, and read out the ends of the chains after evolution under the DQ Hamiltonian. Both these data sets were fitted by the analytical expression (\ref{eq:DQend}).
As it is evident from near perfect fitting, the analytical model explains the experimental data quite precisely.

\begin{figure*}[t]
   \includegraphics{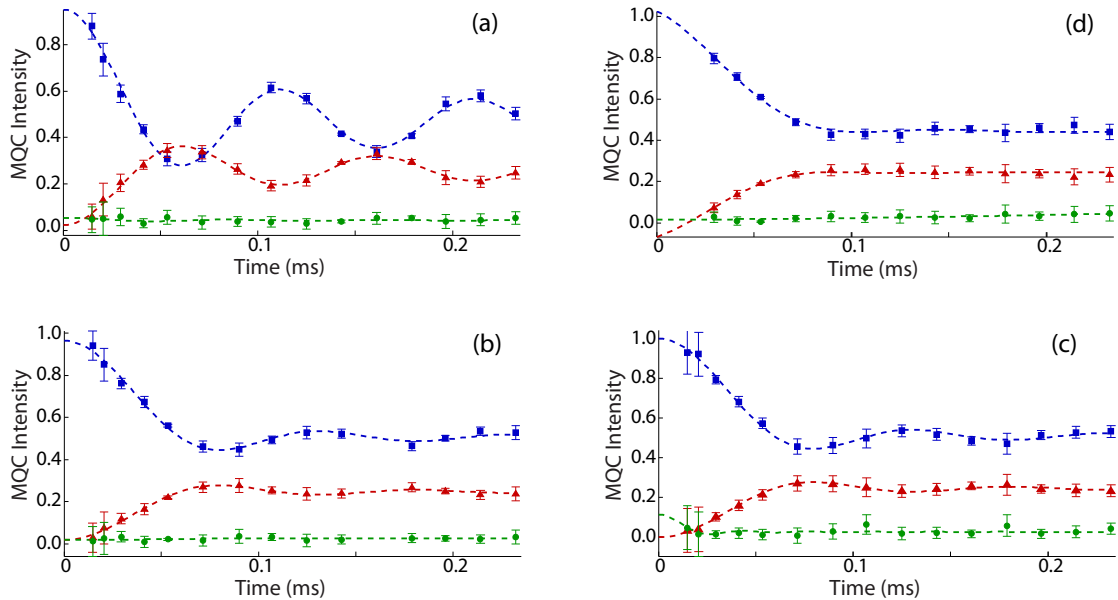}
   \caption{(Color online) Evolution of multiple quantum coherences  (0Q blue squares, 2Q red triangles, 4Q green circles):  (a) Initial state: $\delta\rho_{th}$. Readout: collective magnetization.  (b) Initial state: $\delta\rho_{\textrm{end}}$. Readout: collective magnetization.  (c) Initial state: $\delta\rho_{th}$ Readout: end readout.
      (d) Initial state: $\delta\rho_{\textrm{end}}$ Readout: end readout.
  The experimental data points are fitted by analytical functions (blue, red and green lines) obtained from DQ Hamiltonian with NN couplings (equations \ref{eq:j02th},\ref{eq:j02end},\ref{eq:j02end} and \ref{eq:j02sre} for  figures (a), (b), (c) and (d) respectively).
  The error bars are estimated from the deviation of 1st order quantum coherence from zero. Fitting of the data gives  dipolar coupling: 7.971 (a), 8.077(b), 8.031 (c) and 8.492 (d) $\times10^3$rad/s.} \label{fig:gkMQCall}
 \end{figure*}

Figs. \ref{fig:gkDQall}(a) and (b) show very different chain dynamics for the two initial states (with and without end selection), giving an experimental validation of our initialization method.
Furthermore, the data and fittings for end selection, Fig.~\ref{fig:gkDQall}(b), and end readout measurements, Fig.~\ref{fig:gkDQall}(c),  are very similar. This indicates the robustness of the readout step.
Finally, Fig.~\ref{fig:gkDQall}(d) shows the evolution of the end polarized initial state under the DQ-Hamiltonian, measured using the ``end readout'' scheme. The fitting function used is given in Eq. \ref{eq:DQsre}, which has the same form as the transport of a single excitation in a pure state chain~\cite{Cappellaro11,Christandl05}: this experiment is thus a direct simulation of spin transport.

In all the above  fittings, we used the following fitting parameters: a scalar multiplier, a baseline constant, the nearest-neighbor dipolar coupling and a  shift of the time axis. The shift of the time axis is needed since there is a delay of a few $\mu$s between the end of the multiple pulse sequence (ideally t=0) and the measurement of the signal.
From the fit we obtain a value for the dipolar coupling of 8.161, 8.172, 8.048 and 8.63 $\times10^3$rad/s for the four experiments, respectively.  Except for the last experiment, the other values agree well with those obtained from similar measurements done on a different FAp crystal~\cite{Zhang09} and also with the theoretical value b=$8.17\times10^3$rad/s resulting from the known structure of FAp.

The small discrepancy in the fitting parameter for the  last  experiment, where we are initializing and reading out the chain ends,  can be explained by imperfections of the end-select and readout schemes.
Unfortunately, the phase cycling scheme does not cancel out zero-quantum terms. Thus, residual polarization on spins 2 and N-1 as well as correlated states of the form $\sigma^z_j(\sigma^+_{j-1}\sigma^-_{j+1}+\sigma^-_{j-1}\sigma^+_{j+1})$ contribute to errors,  lowering the fidelity with the desired state. This effect is more important for the last experiment, since  errors in the two selection steps accumulate. Moreover, the end readout scheme works well only if applied to the  ideal  state which is expected after transport.  The deviation of the prepared initial state from the ideal state further contributes to the error in the final measured data. Still, the agreement of the experimental data with the analytical model indicates that these errors are small and do not invalidate the scheme

\subsection{Experimental Results: Multiple Quantum Coherences}
We present the results of the second set of experiments in Fig.~\ref{fig:gkMQCall}, which shows the evolution of the 0, 2nd and 4th order coherence intensities experimentally measured for different initial states and readouts.
Fig.~\ref{fig:gkMQCall}(a) shows the usual MQC signal, obtained measuring the collective magnetization and starting from an  initial thermal state (Eq. \ref{eq:thermalstate}). The data points are fitted by analytical functions (\ref{eq:j02th}), given in appendix~\ref{sec:App_MQCNMR}.  The only variable used in these fittings was the dipolar coupling constant,for which we   obtained  the value  $b=7.971\times10^3$ rad/s.
 The data shown in Fig.  \ref{fig:gkMQCall}(b) has been measured by first selecting the ends of the chain and then performing the MQC measurement where collective magnetization is read out. Fig.  \ref{fig:gkMQCall}(b), on the other hand, shows the data for MQC measurements starting from thermal initial state but reading out only the spins at the chain ends.
 Both these data were fitted by the analytical functions (\ref{eq:j02end}),  giving $b=8.077$ and 8.031 $\times10^3$ rad/s respectively.
 This is in very good agreement with the values obtained from the quantum transport measurements.
\begin{figure}[b!]
 \includegraphics{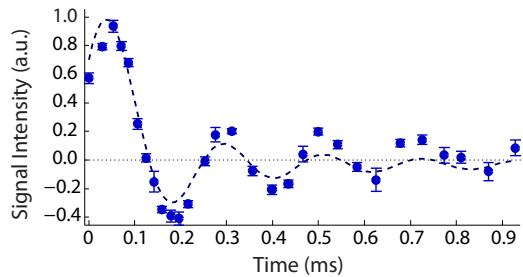}
  \caption{(Color online) Logical initial state and readout of collective magnetization. a) Evolution under DQ Hamiltonian.  Data points are the experimental data and lines are the fits using the analytical model described in appendix A.(equation \ref{eq:DQlog}. Fitting of the data gives  dipolar coupling: 7.551 $\times10^3$rad/s. }
  \label{fig:gklog}
\end{figure}
\begin{figure*}
   \includegraphics{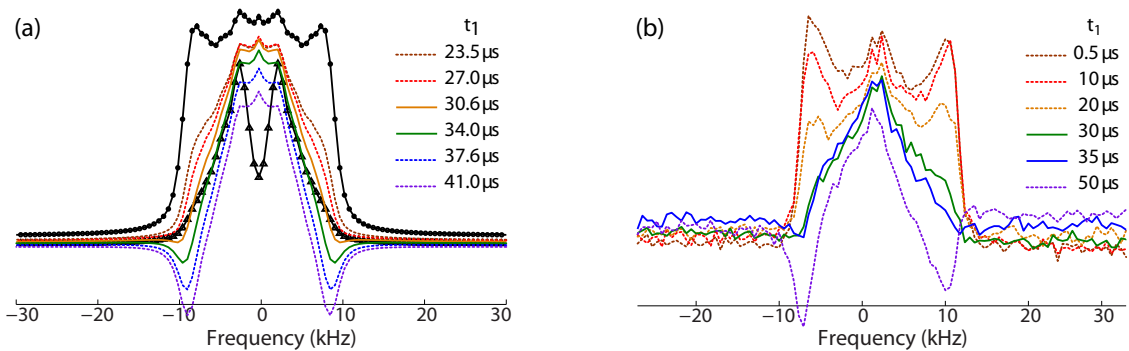}
  \caption{(Color online) (a) Simulated $^{19}$F lineshapes for a chain of 11 spins, for thermal initial state (circles), ideal end-polarized state (triangles) and for an initial state prepared via a simulated~\ref{seq:P1} sequence, with varying $t_1$ times. (b) Experimentally measured $^{19}$F NMR linehsape after state initialization performed with the \ref{seq:P1} sequence for various t$_1$ time. Solid lines are for t$_1=30\mu$s and $35\mu$s, which give the narrowest linewidth and the best state preparation. }\label{fig:gkline}
\end{figure*}
 We remark again that the results obtained from end selection, Fig.  \ref{fig:gkMQCall}(a), and end readout measurements, Fig.  \ref{fig:gkMQCall}(b),  are very similar, thus validating the effectiveness of the readout step.
 The data for the case where we initialize the ends of the chains before letting the system evolve under DQ Hamiltonian and then read out the ends is shown in Fig.~\ref{fig:gkMQCall}(d).  Good fitting of the data  with Eq. (\ref{eq:j02sre}) was obtained for b=8.492$\times10^3$ rad/s and shifting the time axis by 15$\mu$s. As mentioned earlier, we expect the errors in selecting the ends of the chains to add up in the experimental data,  resulting in a slightly higher value of the fitting parameter $b$. The first two data points in the above mentioned figures were measured using a 4 pulse sequence to implement DQ Hamiltonian (instead of a standard 8 pulse sequence), leading to greater error bars for these data points.

\subsection{Quantum Information Transport}
To demonstrate our ability to experimentally simulate not only the transport of classical information, as encoded in the spin polarization, but also of quantum information, we prepared and studied the evolution of one of the logical states in Eq. (\ref{eq:LogDQ}).
The logical state $\delta\rho_y^L$ was prepared by the scheme described in section \ref{sec:EndChain} (similar schemes could be used to prepare the other states in the operator basis). Figure~\ref{fig:gklog} shows the evolution of this state under the DQ Hamiltonian and readout of collective magnetization. The data points were fitted by the expression in Eq.~(\ref{eq:DQlog}) giving the value of dipolar coupling constant as 7.551 $\times10^3$rad/s. We note that the scheme for preparing this logical state involves selecting the ends of the chain and then creating MQC and filtering out all the terms except the double quantum terms. The errors involved in both these steps add up and result in deviation of measured data points from the analytical function. The experimental results, however, follow the expected analytical models and the dipolar coupling constants obtained from these experiments  agree very well with those obtained from other measurements.

In all the above described measurements where ends of the chains were selected and initialized, we used t$_1=30.3\mu$s in the pulse sequence \ref{seq:P1}. As described  above, this value was obtained by  selecting the time when  polarization of the spins at the ends of the chains is non zero while it has  decayed to zero for the other  spins, as a result of evolution under the internal dipolar Hamiltonian.  In order to confirm this value experimentally, we repeated the transport and MQC  measurements at different values of t$_1$ (not shown here). We observed that t$_1$ = $30.3\mu$s gave the best fittings with the analytical functions, pointing towards the fact that the fidelity of end selection is highest for t$_1$ = $30.3\mu$s. This is further confirmed by a detailed $^{19}$F  NMR lineshape analysis as described in the next section.

\subsection{$^{19}$F  NMR lineshape analysis}\label{sec:lineshape}
\begin{figure}[b]
   \includegraphics{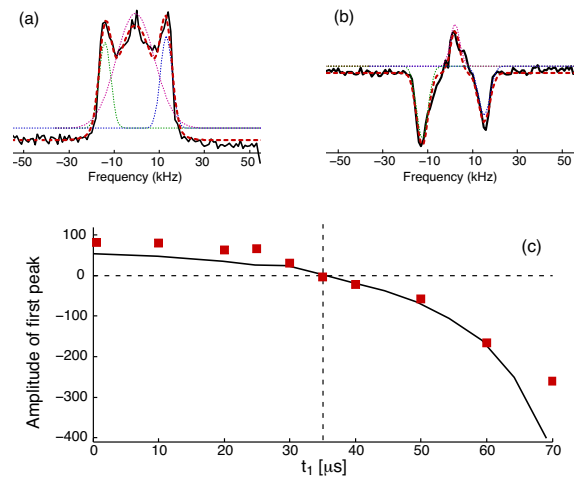}
  \caption{{(Color online) (a)The $^{19}$F NMR lineshape (for t$_1$=10$\mu$s) can be fitted by 3 Gaussian lines. Individual Gaussian lines are shown as thin dotted lines, sum of the lines is thick red. (b) The $^{19}$F NMR lineshape (for t$_1$=60$\mu$s) can be fitted by 3 Gaussian lines out of which 2 lines have negative amplitude. (c) Amplitude of 1st Gaussian line as a function of t$_1$ used to fit the experimental data (red squares) and simulated spectra (black line). }}\label{fig:gkfit1}
\end{figure}

A system comprising  linear chains of spins, such as Fluorapatite, is of immense interest for NMR lineshape calculations by virtue of the simplicity it offers as compared to a 3-dimensional system~\cite{Sur75b,Engelsberg73}. The $^{19}$F spins in FAp have a characteristic 3 peak lineshape, which shows a strong angular dependence~\cite{VanDerLugt64}. In our study, we utilize this angular dependence to align the crystal parallel to the external magnetic field in order to minimize interchain couplings.

\begin{figure}[b]
  \includegraphics{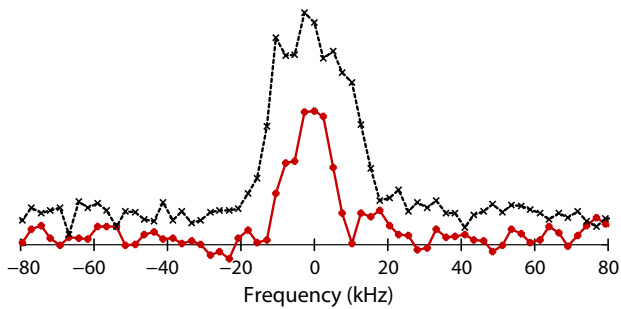}
  \caption{(Color online) $^{19}$F NMR lineshape measured for thermal (black, dotted) and end polarized (red) initial state using pulse sequence \ref{seq:P2}.}\label{fig:gkSEline}
\end{figure}
The measured lineshapes for thermal initial (t$_1$ = 0$\mu$s) and end polarized (t$_1$ = 30.3$\mu$s) using sequence \ref{seq:P2} are shown in Fig. \ref{fig:gkSEline}. Due to long T$_1$ of $^{19}$F nuclei, it was not possible to measure as many points as in the standard FID measurements. This resulted in low resolution of the Fourier transformed signal and hence the three characteristic peaks of FAp lineshape are not resolved properly. Nonetheless, the differences in the lineshape and line width for the two signals are distinctly visible.

The lineshape provides not only signatures of the system dimensionality, but of its initial state as well, thus  we expect to see  qualitative differences in the NMR spectra for the thermal and end polarized states.
We calculated the Free Induction Decay (FID)  for a  chain of  $N=11$ $^{19}$F nuclear spins evolving under the secular dipolar Hamiltonian  (Eq. \ref{eq:dipHam}) with  coupling b = $8.1\times10^3$ rad/s.
These calculations were performed also for end-polarized states, obtained by simulating the sequence \ref{seq:P1} for different values of end selection time (t$_1$). The NMR spectra obtained from these simulations are shown in Fig. \ref{fig:gkline}(a), together with the lineshapes for thermal state (Eq. \ref{eq:thermalstate}) and ideal end polarized state (Eq. \ref{eq:endpolstate}).
A few observations  are worth pointing out.
The NMR linewidth shows a progressive narrowing as t$_1$ is increased starting from t$_1$ = 0$\mu$s (corresponding to the thermal state). The linewidth is narrowest for t$_1$ = 30.6$\mu$s and then starts to increase gradually but now with an anti-phase (dispersive) component.
A simple explanation of these features can be obtained by considering the signal as arising from a competition between the signal of the end-chain spins and the signal from the spins in the bulk.
The spins at the chain extremities are dipolarly coupled to only one spin,  hence we expect a splitting of the NMR line into a doublet. The  spins in the chain, instead, are each coupled to two nearest neighbors and hence produce a broader NMR line, split into a triplet. As the first contribution increases with increasing t$_1$ time, the linewidth narrows. At
even longer times, (t$_1$ beyond 30.6$\mu$s) two-spin correlations are created, that give rise to a dispersive spectrum under the subsequent dipolar Hamiltonian evolution. These anti-phase terms keep increasing for longer t$_1$ times.
We observe that none of the simulated lines exactly replicates the expected lineshape for the ideal end polarized state. This is most likely due to less than 100$\%$ fidelity of end selection and initialization step. However, for t$_1$ = 30.6$\mu$s, the width of the spectrum  is very close to the ideal lineshape.

The simulated lineshapes are however in good agreement  with the corresponding experimentally measured lineshapes (Fig. \ref{fig:gkline}(b)), apart from a slight asymmetry, which might have been introduced by a misalignment of the crystal with respect to the magnetic field. In particular, the experimental spectra for t$_1$ = 30$\mu$s and 35$\mu$s have a width very close to the ideal end selected spectrum.

In order to draw a quantitative comparison between simulated and measured NMR spectra, we fit the lineshapes at different t$_1$ times with a model comprising 3 Gaussian lines, at frequencies shifted by the nearest-neighbor dipolar coupling.
As shown in Fig.~(\ref{fig:gkfit1}), both simulated as well as experimental lineshapes could be fitted reasonably well with this model.
Since the outer lines in the Gaussian model arise only from the spins inside the chains, their amplitude   is expected to show a zero crossing at t$_1$ where the chain extremities have maximum contribution to the measured lineshape and the bulk spins polarization has a very small contribution. In our data, this is seen at 35$\mu$s indicating that end selection has maximum fidelity for this value of t$_1$. This is in close accordance with 30.3$\mu$s obtained from an optimization of the  DQ-Hamiltonian evolution fitting and used for our measurements.

In the lineshape analysis described above, all the NMR measurements were carried out by using a $\pi$/2  pulse and measuring the resultant FID. This scheme reads out the collective magnetization from all the nuclei within the chains. It would be interesting to isolate the signal contribution  from the nuclei located at the chain ends.   This was achieved by means of pulse sequence \ref{seq:P2},
\begin{equation}\label{seq:P2}
    \ref{seq:P1}\text{ --- } \pi/2 \text{ --- } (\tau)  \text{ --- }  \pi/2 \text{ --- }\ref{seq:P1},
     \tag{P2}
     \nonumber
\end{equation}
where the end selection step (sequence \ref{seq:P1}) is used to polarize as well as read out the nuclei at the chain ends. The selection and readout steps are separated by a variable delay ($\tau$) which enables measurement of free induction decay (evolution under the natural dipolar Hamiltonian) as a function of time. The Fourier transform of this FID results in a NMR lineshape where only the nuclei at the chain ends contribute.

\section{Conclusion}
\label{sec:Conclusion}
In conclusion, we have studied  spin transport in linear chain of nuclear spins and experimentally demonstrated  addressability of spins at the ends of the chain by means of NMR control schemes. We have shown that even though NMR implementation allows only collective control and observables, we could achieve initialization as well as readout capabilities through a combination of coherent and incoherent control.
These techniques can be used to prepare state of relevance for quantum information transport as well as to monitor the dynamics of the end-chain spins as it evolves under the DQ Hamiltonian obtained via collective manipulation of the natural dipolar interaction. We validated our method by comparing the experimental results with analytical solutions based on an idealized model, which applies to the timescales explored in the experiment. The good agreement of the data with the analytical formula confirms the preparation and readout of the desired state.

In addition, we probed the states and their evolution by means of multiple quantum coherence measurements, which reveal information
about multi-spin correlations. Again, a good conformity of the experimental results with the theoretical model was observed. We further optimized the end-selection scheme by a detailed analysis of the F$^{19}$ NMR lineshapes obtained from  collective thermal magnetization and the end-polarized state respectively.

Although we cannot achieve universal control of the end-chain spins, the initialization and readout capabilities demonstrated in this work will allow us to experimentally  characterize quantum transport in spin chains. It will enable us, for example, to explore non-idealities that emerge e.g. at longer times from non-nearest neighbor couplings as well as couplings to other chain, and from the interaction of the chains with the environment.
Additionally these methods will allow further experimental studies of control schemes that can enable perfect fidelity transfer.

\vspace{12pt}
\textbf{Acknowledgments}\\
This work was partially funded by NSF under grant DMG-1005926. We are grateful to Prof. Katsuya Teshima  (Shinshu University, Japan) for providing the Fluorapatite crystal used in this work.
It is a pleasure to thank  Lorenza Viola and Chandrasekhar Ramanathan  for fruitful discussions.
\appendix
\section{Analytical solution for the evolution under the DQ-Hamiltonian}
\label{sec:App_Analytical}
Information transport in linear spin chains has been often studies in the literature as arising from the evolution under the isotropic XY Hamiltonian, $\ham_{XY}=\sum_{j <\ell} \frac{b_{j\ell}}{2}
(\sigma^x_j\sigma^x_{\ell}+\sigma^y_j\sigma^y_{\ell})$. For mixed-state chains, we showed that the DQ-Hamiltonian (Eq.~\ref{eq:dqHam}) can as well drive the transport, since  it is connected to XY Hamiltonian  by a similarity transformation.  We can thus simulate quantum transport with an Hamiltonian that (unlike the XY Hamiltonian) can be implemented experimentally using dipolar Hamiltonian and collective RF pulses using a standard sequence~\cite{Yen83,Ramanathan03}.

The  evolution of 1D spin system under a DQ Hamiltonian is exactly solvable in the nearest neighbor (NN) limit (only nearest neighbor couplings are present and all are equal to b), by invoking a Jordan-Wigner mapping onto a system of free fermions~\cite{Cappellaro07a,Yen83,Ramanathan03,Feldman96}. Various formulas describing different initial states and observables have been reported~\cite {Cappellaro07l,Cappellaro11,Ramanathan11} and we reproduce here the formulas  we used to interpret our experimental results in the main text.

The analytical solutions for the evolution of the thermal state and end polarized state, when measuring the collective magnetization, are given by:
\begin{eqnarray}
S^{th}(t) &=& \sum_{p=1}^{N} A_{p,p}(2t). \label{eq:DQ}
\end{eqnarray}
\begin{eqnarray}
S^{\textrm{end}}(t) &=& \sum_{p=1}^{N} A_{1,p}^{2}(t). \label{eq:DQend}
\end{eqnarray}
with
\begin{eqnarray*}
A_{j,q}(t) &&=\sum_{m=0}^{\infty} i^{2m\nu}[i^{\delta}J_{2m\nu+\delta}(2bt)-i^{\sigma}J_{2m\nu+\sigma}(2bt)] \nonumber \\
& &+ \sum_{m=1}^{\infty} i^{2m\nu}[i^{-\delta}J_{2m\nu-\delta}(2bt)-i^{-\sigma}J_{2m\nu-\sigma}(2bt)]
\end{eqnarray*}
Transport from one end of the chain to the other is described by
\begin{eqnarray}
S^{sre}(t) &=& A_{1,1}^{2}(t)+A_{1,N}^{2}(t),\label{eq:DQsre}
\end{eqnarray}
which corresponds to the experimental situation where we prepared the end-polarized state and measured the chain ends only.
Finally, we can derive the expected signal arising from the collective magnetization when the initial state is the logical state $\delta\rho_{yL}$:
\begin{eqnarray}
S^{yL}(t) &=& A_{1,2}(2t)+A_{N-1,N}(2t). \label{eq:DQlog}
\end{eqnarray}

\section{Multiple Quantum NMR Spectroscopy}
\label{sec:App_MQCNMR}

Multiple  Quantum  spectroscopy is a powerful technique in  Nuclear Magnetic Resonance. It has the ability to simplify complex spectra by revealing some of the forbidden transitions.  Additionally, creation and evolution of quantum coherences can be used to probe the dynamics of a correlated many-spin system giving insight into dimensionality of spin system, distribution of coupling constants and effects of motions and quantum transport~\cite{,Feldman97,Doronin00,Baum85,Levy92,Ramanathan03,Cho05,Cho06}. Multiple  Quantum  Coherences (MQC) can be excited by driving the spin system by irradiation with cycles of multiple pulse sequences consisting of RF pulses and delays.
For example,  the so-called Double Quantum Hamiltonian (Eq.~\ref{eq:dqHam}) creates even quantum coherences from the longitudinal magnetization.
 Standard NMR techniques, however, measure only single quantum coherences. Thus, in order to probe the spin dynamics, it is necessary to indirectly encode the signature of the dynamics into single quantum coherences which can be directly measured inductively. This is done by labeling each coherence order with a different phase: If $\rho_i$ is the initial density matrix,  the final density matrix $\rho_f$ is given by:
\begin{equation*}
\rho_f = U_{MQ}^\dag U_{\phi} U_{MQ} \rho_i U_{MQ}^{\dag}
U_{\phi}^\dag U_{MQ},
\end{equation*}
where $U_{MQ} = \exp(-i \bar{\cal H}_{DQ}t)$, and $U_{\phi} = \exp(-i\phi\sigma_z/2)$  is a rotation about the z axis by an angle $\phi$. Under this rotation, a coherence term of order $n$ will acquire a phase $n\phi$. In order to extract the information about the distribution of MQC, each measurement must be repeated while incrementing $\phi$ from 0 to 2$\pi$ in steps of  $\delta\phi = 2\pi/2K$ where $K$ is the highest order of MQC we wish to encode. A $\pi$/2 pulse is used to read out the signal at the end of the experiment. Finally, Fourier transform of the signal with respect to $\phi$  yields the coherence order intensity:
\begin{eqnarray*}
J_n(t) &=& \sum_{k=1}^{K} S_z^k(t) e^{-ikn\delta\phi},
\end{eqnarray*}
where  $S_z^k(t)=\tr{\rho_f^k(t)\sigma_z}$ is the signal acquired in the $k$th measurement.
The  technique of MQC spectroscopy outlined above is particularly well suited to study  information transport by means of DQ Hamiltonian, since the coherence intensities show characteristics signatures of the occurred transport~\cite{Cappellaro07l}.

 In a 1D system with NN coupling, only zero and double quantum coherence orders are created by the DQ Hamiltonian. Starting from the thermal initial state, the normalized intensities of the zero and DQ coherences predicted by the analytical model are given by:
\begin{align}
\begin{array}{l}
\displaystyle J_0^{th}(t) = \frac1{N} \sum_k \cos^2[4bt\cos(\psi_k)],  \\
\displaystyle J_2^{th}(t) = \frac1{2N} \sum_k \sin^2[4bt\cos(\psi_k)],
\end{array}
\label{eq:j02th}
\end{align}
where as before $N$ is the number of spins in the chain and
$\psi_k=k\pi/(N+1)$.

For the end-select state, zero and double quantum coherence intensities given by the analytical model are as follows:
  \begin{equation}
  \begin{array}{l}
  \displaystyle
  J_0^{\textrm{end}}(t) = \frac{2}{N+1} \sum_k \sin^2(\psi_k)  \cos^2[4bt\cos(\psi_k)],  \\
   \displaystyle J_2^{\textrm{end}}(t) = \frac1{N+1} \sum_k \sin^2(\psi_k)  \sin^2[4bt\cos(\psi_k)].
  \end{array}
  \label{eq:j02end}
  \end{equation}
  In both Eqs. (\ref{eq:j02th}) and (\ref{eq:j02end}), the normalization is
  chosen such that $J_0+2J_2=1$.

In the case where the chain end selection is operated for both initialization and readout, the  normalized zero and double quantum intensities are given by~\cite{Cappellaro07l}:
\begin{align}\label{eq:j02sre}
J_0^{sre}(t)=& \frac{4}{(N+1)^2}\displaystyle \sum_{k,h} \sin^2(k)\sin^2(h)\cos^2(\psi_k+\psi_h) \nonumber\\
&\times(1+\cos[Nk+k]\cos[Nh+h]),  \\
 J_2^{sre}(t) =&\frac{2}{(N+1)^2} \displaystyle \sum_{k,h} \sin^2(k)\sin^2(h)\sin^2(\psi_k+\psi_h) \nonumber\\
&\times(1+\cos[Nk+k]\cos[Nh+h]),  \nonumber
\end{align}
where the superscript `$sre$' refers to select and read ends.

Finally, for the logical state $\delta\rho_{yL}$, the MQC intensities are as follows:

\begin{equation}
  \begin{array}{l}
  \displaystyle
  J_0^{yL}(t)=\frac{2}{N+1}\sum_{k=1}^N\sin(\psi_k)\sin(2\psi_k)\sin\left(8bt\cos\psi_k\right),  \\
   \displaystyle J_2^{yL}(t)=\frac{-1}{N+1}\sum_{k=1}^N\sin(\psi_k)\sin(2\psi_k)\sin\left(8bt\cos\psi_k\right).
  \end{array}
  \label{eq:j02log}
  \end{equation}

\bibliographystyle{apsrev4}

%

\end{document}